\pdfoutput=1

\documentclass[aps,prl,twocolumn,superscriptaddress,nobibnotes]{revtex4-2}

\usepackage[utf8]{inputenc}
\usepackage{bm}
\usepackage{siunitx}

\usepackage{enumerate}
\usepackage{amsfonts}
\usepackage{amsmath}
\usepackage{amssymb}
\usepackage{color}
\usepackage{soul}
\usepackage{float}
\usepackage{bm}
\usepackage{graphicx}
\usepackage[colorlinks,allcolors=blue]{hyperref}
 
\let\phi=\varphi
\let\epsilon=\varepsilon

\newcommand{\rv}[0]{\bm{r}}

\definecolor{DarkRed}{rgb}{0.80,0,0}
\definecolor{DarkGray}{rgb}{0.7,0.7,0.7}

\usepackage[capitalise]{cleveref}

\begin{document}
\title{\bf Grain-boundary 
topological superconductor}
\author{Morten Amundsen}
\affiliation{Nordita, KTH Royal Institute of Technology and Stockholm University,
Hannes Alfvéns väg 12, SE-106 91 Stockholm, Sweden}
\author{Vladimir Juri\v{c}i\'c}\email{ Electronic address: juricic@nordita.org}
\affiliation{Nordita, KTH Royal Institute of Technology and Stockholm University,
Hannes Alfvéns väg 12, SE-106 91 Stockholm, Sweden}
\affiliation{Departamento de F\'isica, Universidad T\'ecnica Federico Santa Mar\'ia, Casilla 110, Valpara\'iso, Chile}

%%%%%%%%%%%%%%%%%%%%%%%%%%%
%%%%%%%%%%%%%%%%%%%%%%%%%%

\maketitle

%-------------------------------------------------------------------------------%
%                                 MAIN ARTICLE                                  %
%-------------------------------------------------------------------------------%

\noindent
{\bf Abstract}\\
{\bf
Majorana zero modes (MZMs) are of central importance for modern condensed matter physics and quantum information due their non-Abelian nature, which thereby offers the possibility of realizing topological quantum bits. We here   show that a grain boundary (GB) defect  can host  a topological superconductor (SC), with a pair of cohabitating MZMs at its end when immersed in a parent two-dimensional gapped topological SC with the Fermi surface enclosing a nonzero momentum. The essence of our proposal lies in the magnetic-field driven hybridization of the localized MZMs at the elementary blocks of the GB defect, the single lattice dislocations, due to the MZM spin being locked to the Burgers vector. Indeed, as we show through numerical and analytical calculations, the GB topological SC with two localized MZMs emerges in a finite range of both the angle and magnitude of the external magnetic field.
Our work demonstrates the possibility of  defect-based platforms for quantum information technology and opens up a route for their systematic search in future.}
\\

%%%%%%%%%%%%%%%%%%%%%%%%%%%%
%%%%%%%%%%%%%%%%%%%%%%%%%

\noindent 
{\bf Introduction}\\
Topological superconductors (SCs) occupy a rather special place in the landscape of topological states of quantum matter since they can feature  Majorana zero modes (MZMs)~\cite{Read-Green-2000,Kitaev2001,Ivanov-2001,Leijnse-NJP2012,beenakker-2013,Sato_2017} which exhibit non-Abelian statistics, and are promising platforms for quantum computing~\cite{Kitaev-003,DasSarma-RMP2008}. A simplest platform for realizing  these exotic quasiparticles is given by the Kitaev model on a one-dimensional (1D) chain of spinless fermions, where they appear at its  ends when $p-$wave pairing prevails~\cite{Kitaev2001}.  %%%%%%%%%%%%%%%%%%%%%%%%%%%%%%%%%%%%%%%%%%%%%%%%%%%%
%%%%%%%%%%%%%%%%%%%%%%%%%%%%%%%%%%%%%%%%%%%%%%%%%%%%
\begin{figure}[t!]
\includegraphics{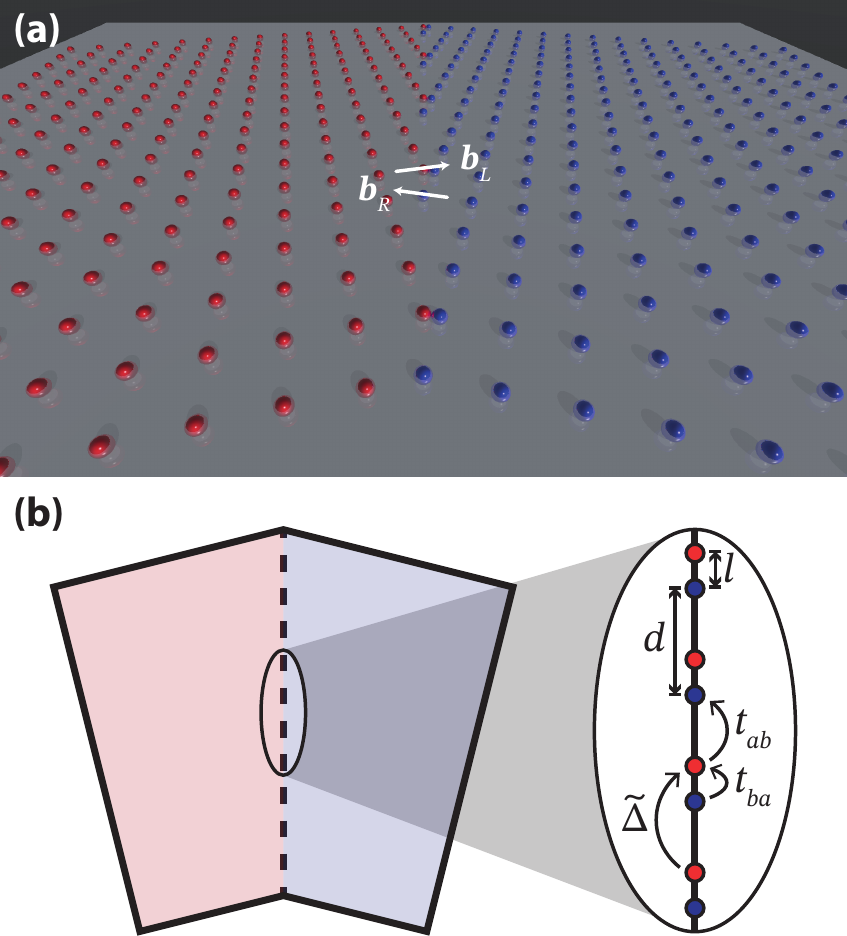}
\caption{{\bf The parent topological superconductor with the grain-boundary defect.} (a) Two misaligned crystalline lattices of a $p-$wave superconductor  form a grain boundary consisting of edge dislocations at their intersection. $\bm{b}_L$ and $\bm{b}_R$ are the Burgers vectors of the two inequivalent edge dislocations, belonging to the left and right sublattice, respectively. (b) Schematic illustration of the effective 1D superlattice formed by the single dislocation defects along the grain boundary, together with the parameters in the effective Hamiltonian~\eqref{eq:effHam}. The parameter $d$ ($l$) indicates the distance between neighboring lattice sites within the same (opposite) sublattice.  The competition between the inter-unit cell $t_{ba}$ and the intra-unit  cell $t_{ab}$ pairings drives the topology of the emergent superconductor along the grain boundary, while $\tilde{\Delta}$ is the next-nearest neighbor intra-unit cell pairing.  }
\label{fig:model}
\end{figure}
%%%%%%%%%%%%%%%%%%%%%%%%%%%%%%%%%%%%%%%%%%%%%%
%%%%%%%%%%%%%%%%%%%%%%%%%%%%%%%%%%%%%%%%%%%%
%%%%%%%%%%%%%%%%%%%%%%%%%%%%%%%%%%%%%%%%%%%%
In turn, this yielded a surge of theoretical proposals in which  Majorana modes emerge through hybridization of the states localized at building blocks (sites) in different  1D  chain-like architectures, constituted by magnetic atoms~\cite{Perge-PRB2013,Pientka-PRB2013} and quantum dots~\cite{Sau-NatComm2012,Leijnse-PRB2012},  with the signatures of the MZMs  reported in experiments~\cite{Perge-Science2013,Ruby-2015,Pawlak2016,Jeon-Science2017,Kim-SciAdv2018,kouwenhoven2022}. However, these platforms are 
highly sensitive to the microscopic details of the system and require fine tuning.

%%%%%%%%%%%%%%%%%%%%%%%%%%
%%%%%%%%%%%%%%%%%%%%%%%%%%
\begin{figure*}
\includegraphics{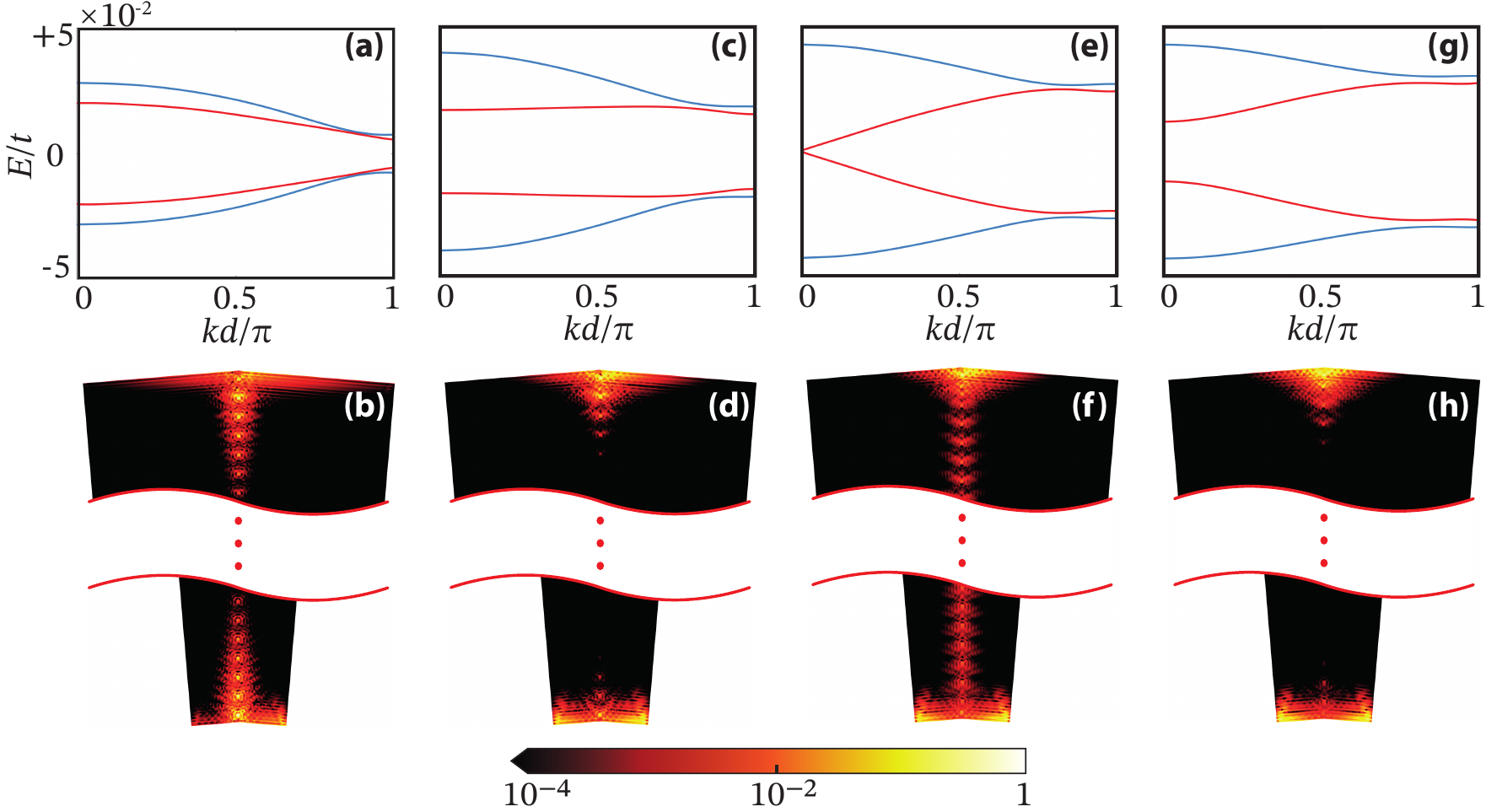}
\caption{{\bf Nature of grain-boundary superconductor}. The band structure [(a), (c), (e) and (g)] and the accompanying local density of states (LDOS) at zero energy [(b), (d), (f), and (h)] for increasing exchange field ($h$) directed along the grain boundary.  Each column corresponds to a particular value of the exchange field, with $h = 0.1t$ in (a) and (b), $h = 0.2t$ in (c) and (d), $h = 0.25t$ in (e) and (f), and $h = 0.28t$ in (g) and (h). Here, $t$ is the hopping parameter.  The two zero-energy modes  localized at the top and the bottom edge in panels (b) and (d) pertain to the parent topological superconductor (SC), see also \cref{fig:spectrum}(a). We note that in figures (e) and (f), the grain boundary is gapless, and  connects two edge zero modes of the parent gapped SC. In panel (h), there are four zero modes, as the grain boundary SC becomes topological, see also~\cref{fig:spectrum}(b) and \cref{fig:bsh}(a). The red and blue coloring in the band-structure plots act only as a visual aid, to more clearly distinguish the bands. To make the localized states at the top and bottom of the grain boundary more visible, the middle part of the LDOS plots has been removed. Here, we set the order parameter to  $\Delta_p = 0.75t$, the inter-sublattice distance between edge dislocations to $d = 12a$, and the slip between sublattices to $l = 5d/12$ in the parent SC Hamiltonian~\eqref{eq:model}. Additional plots showing the entire geometry are available in Supplementary Note~2, see Supplementary Figure 3.  }
\label{fig:bsmu}
\end{figure*}
%%%%%%%%%%%%%%%%%%%%%%%%%%%%
%%%%%%%%%%%%%%%%%%%%%%%%%%%%%%%

In this work, we identify an alternative to these proposals, the  grain boundary (GB) defects, which are at low angles built of an array of edge dislocations, forming due to the accumulated stress at the interfaces between the crystalline grains~\cite{Sutton1995,Hilgenkamp2002,Han2018}. Such extended defects may accommodate self-organized semimetals in parent static topological insulators~\cite{Slager-PRB2016},   experimentally observed in  
1T'-MoTe$_2$~\cite{Kim2020}, and were also recently proposed in dynamic topological crystals~\cite{Salib-2022}. The elementary building blocks of the GBs, the single dislocations, as it is by now well established, can host topologically and symmetry protected modes~\cite{Ran-NatPhys2009,Teo-PRB2010,Juricic-PRL2012,Asahi-2012,Slager-PRB2014,HughesYao2014,Roy-Juricic-PRR2021,Nag2021,hu2022dislocation}, and are experimentally observed in bulk topological crystals~\cite{Hamasaki-APL2017,Nayak-SciAdv2019}. A mechanism for their emergence, operative in both insulators and SCs, is the dislocation-magnetic-flux correspondence:   a lattice dislocation, characterized by   a Burgers vector ${\bf b}$, sources an effective magnetic flux, $\Phi={\bf K}\cdot{\bf b}\, ({\rm mod}\, 2\pi)$, which topologically frustrates the bulk quasiparticles emerging at a momentum ${\bf K}$ in the Brillouin zone (BZ)~\cite{Ran-NatPhys2009,Teo-PRB2010,Juricic-PRL2012,Asahi-2012}. In a two-dimensional (2D) topological SC, dislocation MZMs therefore may appear  when the Fermi surface encloses a non-$\Gamma$ point in the BZ, as a consequence of a nontrivial flux sourced by a dislocation~\cite{Asahi-2012}. This class of topological SCs is in the following referred as translationally-active, analogous to their insulating counterparts~\cite{Juricic-PRL2012,slager-natphys2013}.  A GB defect in  a translationally-active SC therefore naturally provides a 1D lattice structure with its elementary constituents hosting the MZMs, which, in turn, hybridize and can possibly yield a topological SC. We emphasize that the effective flux carried by edge dislocations  $\Phi = \pi$, thus resembling a half-quantum vortex. The behavior of a grain boundary is therefore distinguished from that of an array of conventional vortices, with the confined flux being an integer multiple of $2\pi$.

\begin{figure}[t!]
\includegraphics[width=\columnwidth]{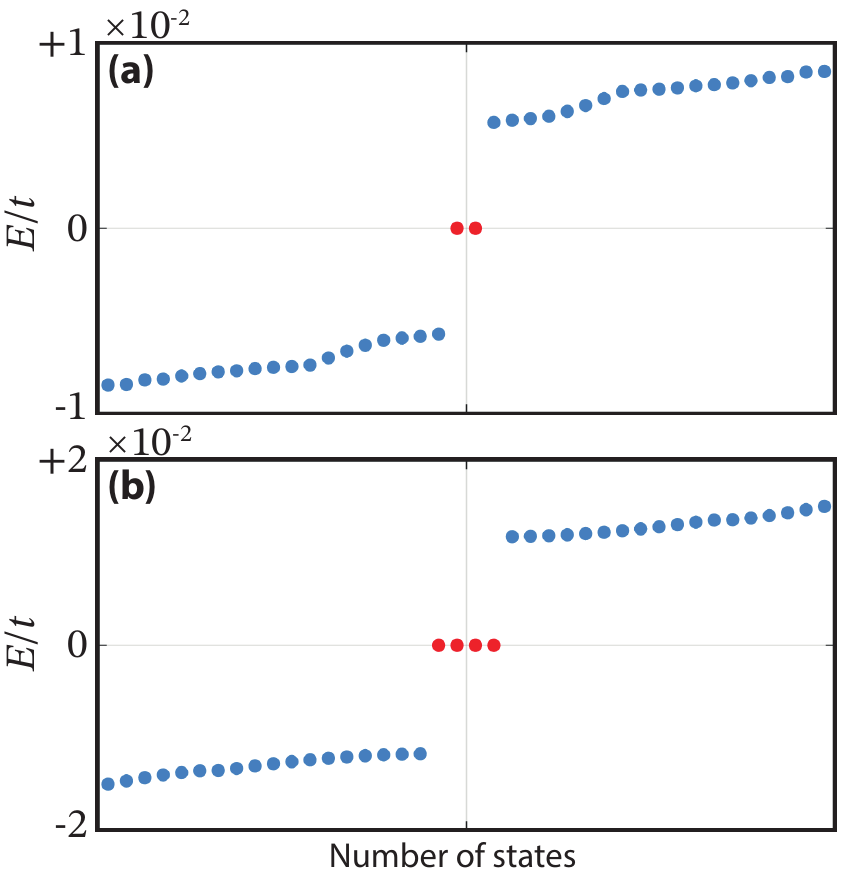}
\caption{{\bf Closest-to-zero energy states of the Majorana  band structure.}  (a) $h/t = 0.1$ and (b) $h/t = 0.28$, with $h$ the exchange field and $t$ the hopping parameter, showing two and four zero-energy modes in the trivial and the topological phase of the grain-boundary superconductor, respectively. These two plots correspond to the zero-energy modes shown in Figs.~\ref{fig:bsmu}(d) and (h), respectively. Two extra zero modes in panel (b) correspond to the grain-boundary topological superconductor, see also~\cref{fig:bsh}. }
\label{fig:spectrum}
\end{figure}

\begin{figure}[t!]
\includegraphics{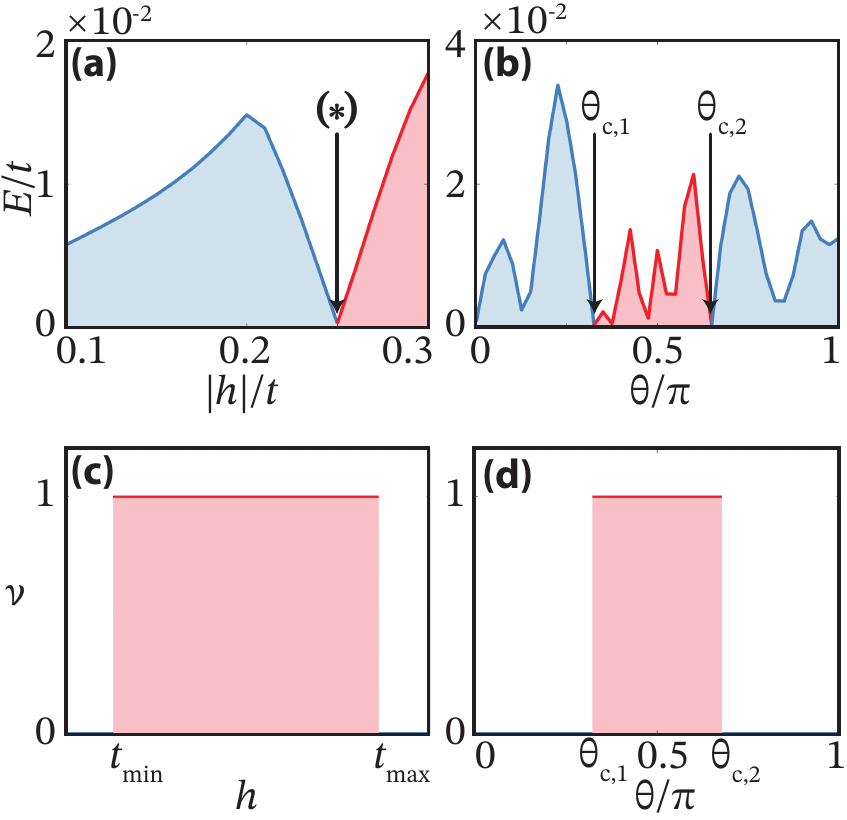}
\caption{{\bf Topological phases of the grain-boundary (GB) Majorana chain.}  Panel (a) shows  numerically calculated minimum gap of the GB band structure as a function of the exchange field magnitude (for $\theta = \pi/2$). Panel (b) shows numerically calculated minimum gap of the GB band structure as a function of   in-plane orientation of the field with respect to the horizontal axis $\theta$ (for fixed $|h|/t = 0.3$, with $h$ the exchange field and $t$ the hopping parameter). The annotations $(*)$ and $\theta_{c,i}$ indicate transitions between the two topologically distinct regimes:  (red) GB chain features two topological Majorana zero modes, and (blue) no endpoint  modes; no parent zero modes are taken into account here. At the point $(*)$ the effective spin splitting produced by the exchange field becomes equal to $t_{\text{min}}$ (the minimum gap in the absence of spin splitting). A reentrance into the trivial regime is expected when the spin splitting becomes equal to $t_{\text{max}}$, which is beyond the limit for which bulk superconductivity survives. The topological invariant $\nu$ of the effective  model in Eq.~\eqref{eq:effHam} as a function of exchange field magnitude, for $\theta = \pi/2$ fixed [panel (c)], and for different  orientations of the field [panel (d)],  for $h\simeq0.3t\simeq 1.25 t_{\text{min}}$ fixed. Comparison between the results from the numerical analysis  and the effective model shows a correspondence between the appearance of GB Majorana zero modes and a nonzero topological invariant, thereby corroborating their topological nature.  In the numerical calculation, the order parameter is set to $\Delta_p = 0.75t$, the chemical potential $\mu = 5.85t$, and the slip between sublattices  $l = 5d/12$.}
\label{fig:bsh}
\end{figure}

We here show that the extended 1D GB defect (Fig.~\ref{fig:model}), as a separate subsystem, can host a topological SC with  a pair of MZMs at its end, when immersed in a parent 2D  translationally-active topological SC, as shown in Figs.~\ref{fig:bsmu} and~\ref{fig:spectrum}.  Such GB topological SC, as we demonstrate, then emerges in a finite range of both the angle and the magnitude of the applied exchange magnetic field, as displayed in Fig.~\ref{fig:bsh}, and corroborated by an effective model, see Eq.~\eqref{eq:effHam}. The GB MZMs are localized at the top of the defect, and, most importantly, are   spatially separated from the edge modes arising from the topologically nontrivial host, see Fig.~\ref{fig:bsmu}.  As we explicitly show, these GB  modes are also protected by an  antiunitary particle-hole-like symmetry of the effective GB Hamiltonian. Finally,  we argue that thin films of Fe-based SCs, doped topological crystalline insulators, such as In-doped SnTe, and designer materials represent  promising platforms for the realization of the proposed emergent GB SC. 

\noindent\\
{\bf Results and Discussion}\\

{\bf Lattice model with the GB defect.} We study a topological $p-$wave SC on a square lattice with lattice constant $a$. Furthermore, this lattice features  a GB defect formed by two regions with the two identical lattices, rotated by an angle $2\alpha$ and slightly shifted by a distance $l$ relative to one another. The geometry of the defect is shown in Fig.~\ref{fig:model}(a). At low angles $\alpha\lesssim15^\circ$, this defect consists of an array of edge dislocations forming a 1D superlattice  with a two-atom  unit cell. The lattice constant is $d$ and  $l$ is the distance between the atoms in the unit cell, as illustrated in Fig.~\ref{fig:model}(b). The opening angle $\alpha$ is related to the dislocation Burgers vector $\mathbf{b}$, as $\sin\alpha=|\mathbf{b}|/d$. See also Supplementary Note 1 and Supplementary Figure 1.

The Bogoliubov-de Gennes  Hamiltonian of the parent SC is $\mathcal{H}=\sum_{{\bf k}}\Psi_{{\bf k}}^\dagger H({\bf k})\Psi({\bf k})$, with $\Psi_{{\bf k}}=(c_{k,\uparrow},c_{k,\downarrow},
c_{-k,\uparrow}^\dagger,c_{-k,\downarrow}^\dagger)^\top$, where  $c_{k,\uparrow}$ ($c_{k,\downarrow}^\dagger$) is the
annihilation (creation) operator for the quasiparticle with
spin up (down) and momentum $\mathbf{k}$, and
\begin{align}
H &= \left[2t(2-\cos k_xa-\cos k_ya)-\mu\right]\sigma_0\tau_3 \nonumber \\
&+ \Delta_p\left(-\sin k_xa\,\sigma_0\tau_2 +\sin k_ya\,\sigma_3\tau_1\right) - \bm{h}\cdot\bm{\sigma}\tau_3.
\label{eq:model}
\end{align}
Here, $(\sigma_0,\bm{\sigma})$ [$(\tau_0,\bm{\tau})$],  are the standard Pauli matrices acting in the spin (Nambu) space, $t$ is the overall energy scale (related to the effective mass of the fermionic quasiparticles), $\mu>0$ is the effective chemical potential, and $\Delta_p$ is the  $p-$wave pairing amplitude~\cite{Asahi-2012}. We also include an in-plane exchange magnetic field,  $\bm{h}$. For $\bm{h}=0$, the parent SC is time-reversal symmetric, characterized by a $Z_2$ topological index, in the class DIII of the tenfold periodic table~\cite{Ryu_2010},  featuring a Kramers pair of gapless Majorana edge states in  
the topological phase.  On the other hand, a finite exchange field, ${\bm h}\neq0$, can partially gap out the edges, yielding the Majorana zero-energy corner modes in a polygonal geometry, thereby changing the parent SC's topology to second order~\cite{Zhu-PRB2018}. The details of the model are given in Supplementary Note~1. Most importantly,  for the values of the parameter $4<\mu/t<8$ and ${\bf h}=0$, the parent SC is fully gapped in the bulk,  topologically nontrivial and encloses the Fermi momentum at  either the $X$ point [$(\pi,0)$] and the $Y$ point [$(0,\pi)$], or the $M$ point [$(\pi,\pi)$] in the BZ~\cite{Asahi-2012}. Therefore,  a single dislocation defect sources a $\pi$ flux, and  binds a pair of the MZMs in this translationally-active SC. These modes in turn make a Wannier basis for a  self-organized  topological SC, which emerges along the GB defect when the exchange field is applied, as we show in the following.

%%%%%%%%%%%%%%%%%%%%%%%%%%%%%%
%%%%%%%%%%%%%%%%%%%%%%%%%
%%%%%%%%%%%%%%%%%%%%%%%%%%%%%%%%
%%%%%%%%%%%%%%%%%%%%%%%%%%%%%%%%%%

{\bf Numerical results.} We now numerically diagonalize  the Hamiltonian in \cref{eq:model}, after implementing it on a real-space square lattice  with the GB defect, shown in Fig.~\ref{fig:model}, see   Supplementary Note~1 for further details. The numerical calculations were performed using the Kwant code~\cite{kwant}. 
To obtain the band structure along the GB superlattice, we  Fourier transform the corresponding eigenstates with respect to the superlattice periodicity ($d$), and identify the peak in the Fourier spectrum at the representative momentum. We consider a model in which the distance between edge dislocations on the same sublattice, $d = 12a$, and introduce a small slip of $l = 5d/12$ between the two sublattices,  which therefore alternate with  spacing of $5a$ and $7a$, and $\alpha=4.8^\circ$. As such, the slip breaks the intra-unit-cell mirror symmetry, increasing the richness of behaviors exhibited by the emergent GB SC. Each sublattice has  width of 30 lattice sites at its bottom edge, and the GB consists of 138 edge dislocations.

%%%%%%%%%%%%%%%%%%%%%%%%%%%%%%%%%
%%%%%%%%%%%%%%%%%%%%%%%%%%%%%%%%
%%%%%%%%%%%%%%%%%%%%%%%%%%%%%%%%%%
%%%%%%%%%%%%%%%%%%%%%%%%%%%%%%%%%%

Indeed, as shown numerically in Supplementary Note 2 and Supplementary Figure 2, in the absence of the exchange field, equidistant edge dislocations will always feature a node in the band structure at the $M$ point. With the introduction of a slip, the grain boundary band structure oscillates between having a node at the $\Gamma$ and $M$ points as a function of the chemical potential, being fully gapped in between. The band structure then features two Kramers pairs of the bands, which are consequence of the particle-hole and time-reversal symmetries, and is topologically trivial.  

To lift the spin degeneracy of the GB band structure, we now apply an in-plane magnetic (exchange) field which is directed along the GB direction. As expected, this yields four nondegenerate bands and opens up a gap at the $M$ point of the GB BZ, as displayed  in \cref{fig:bsmu}, where the GB band structure is shown as the magnitude of the exchange field increases, together  with corresponding local density of states, see also Supplementary Figure 3. We point out that the four bands never cross, but instead feature anti-crossings with a rather small gap. 
In Figs.~\ref{fig:bsmu}(a) and (b), with $h/t = 0.1$, there are two zero-energy modes [see also Fig.~\ref{fig:spectrum}(a)] at the top and bottom edges, which are a consequence of a nontrivial topology of the parent SC. 
As the exchange field is further increased, $h/t = 0.2$, Figs.~\ref{fig:bsmu}(c) and (d), the GB gap becomes larger, and more importantly, it yields a nearly perfect flat band, as the  system approaches the band gap closing at the $\Gamma$ point.  This gap closes for $h/t = 0.25$, as shown in Figs.~\ref{fig:bsmu}(e) and (f), and the GB Majorana band structure is gapless. On the other hand, the parent bulk state remains gapped and topological, as can be directly seen from the well localized zero modes at the top and the bottom edges connected through the gapless GB. 
The gap opens  at the $\Gamma$ point for a stronger exchange field,  $h/t = 0.28$, as displayed in Figs.~\ref{fig:bsmu}(g) and (h). However, there are  now {four} zero energy modes in the spectrum, as shown in Fig.~\ref{fig:spectrum}(b). Two of these states are localized at the two bottom corners, and represent the zero-energy edge states of the parent SC. Most importantly, the two remaining zero modes appearing at the top of the grain boundary are a direct consequence of a topologically nontrivial superconducting state emerging along the GB, as we also further argue below by numerical means, and show analytically within an effective model for the GB band structure.

To corroborate the topological origin of the two GB zero modes, we first notice that their emergence coincides exactly with the GB gap closing, as highlighted in \cref{fig:bsh}(a), and shows the GB gap  as a function of the magnitude of the exchange field, directed along the GB. In the blue region, there are two  zero-energy edge modes present, as long as  the GB SC is fully gapped. On the other hand, there are four modes in the red region, two of which are separated from the ones at lower exchange field (blue region) by the gap closing at the $\Gamma$ point of the GB BZ. 

Additional control over  the GB topological SC  can be achieved by rotating the exchange field, as we show for  $h/t = 0.3$. We find that when the exchange field is orthogonal to the grain boundary ($\theta=0$), the GB features a topologically trivial SC, as seen in \cref{fig:bsh}(b). As the angle $\theta$ is increased anticlockwise towards a parallel orientation ($\theta = \pi/2$), we observe that the GB gap closes at an angle of $\theta_{c,1}\simeq58.5^\circ=1.02$ radians, and the two endpoint GB zero modes emerge upon the gap reopening. Further rotation of the exchange field tunes the GB back to the topologically trivial phase, at the critical angle $\theta_{c,2}\simeq117^\circ=2.04$ radians, where the two GB modes concomitantly disappear.  These features  strongly suggest that the GB SC undergoes a re-entrant  topological phase transition from the topologically trivial to the nontrivial regime and back. This behavior is captured within the effective model for the GB SC, as we demonstrate in the following.

{\bf Effective model.} Nontrivial topology of the GB superconducting state can be accounted for within a rather simple effective  model obtained by projecting the parent Hamiltonian on the subspace formed by the Majorana modes at the dislocation defects forming the GB superlattice, see Supplementary Notes~3 and 4.  The GB consists of edge dislocations with alternating Burgers vectors between the left and the right lattice, see Fig.~\ref{fig:model}(a), and Supplementary Note~1. We  can therefore treat the defect as a 1D lattice with a two-atom basis, as illustrated in \cref{fig:model}(b). Crucially, in contrast to an edge that features spin locked parallel to it, the spins of the localized states at edge dislocations are, as shown in Supplementary Note~3, locked to the Burgers vector, which for the left (right) sublattice, is inclined by an angle of $\alpha$ ($\pi-\alpha$). Furthermore, the form of the single-dislocation modes, found in Supplementary Note~3, implies that the overlap integrals of two edge-dislocation modes located at positions $\rv_i$ and $\rv_j$  behave as  $t_{ij} \sim e^{- r_{ij}/r_0}\cos \beta r_{ij}$, with $r_{ij} = |\rv_i-\rv_j|$. Here, the parameter $r_0$ is the inverse localization length, typically of the order of a few lattice constants, and is a function of the pairing amplitude $\Delta_p$, whereas the oscillation period $\beta$ is  dominated by the Fermi momentum. Therefore, in an effective model for a GB with a sparse array of dislocations, it should be enough to keep the hoppings only between the nearest neighbors. 

From the GB geometry,  shown in \cref{fig:model}(b), by projecting the parent Hamiltonian in Eq.~\eqref{eq:model} on the Wannier basis of the single-dislocation Majorana modes localized on the two  sublattices,  we obtain an effective tight-binding Hamiltonian $\mathcal{H}_{{\rm eff},0}=\sum_{k}\psi^\dagger H_{{\rm eff},0}\psi$, with $\psi_k=(a_{k,\uparrow},b_{k,\uparrow},a_{-k,\downarrow},b_{-k,\downarrow})^\top$, see Supplementary Note~4  for technical details,
\begin{align}\label{eq:effHam}
    H_{{\rm eff}} = \begin{pmatrix} H_0(k) & P(k)  \\  P^\dagger(k) & -H^\top_0(-k)  \end{pmatrix},
\end{align}
where 
\begin{align}
H_{0}(k)& = \begin{pmatrix} 2\tilde{\Delta}\sin kd & -iW(k) \\ i W^*(k) & -2\tilde{\Delta}\sin kd  \end{pmatrix},
\end{align}
while the effective pairing term is induced purely  by the exchange field
\begin{align}
P(k)& = ih\begin{pmatrix} \sin(\alpha-\theta) & \tilde{V}(k) \\  \tilde{V}^*(k) & \sin(\alpha+\theta)  \end{pmatrix}.
\end{align}
Here,  $W(k) = e^{ikl}\left(t_{ab} + t_{ba}e^{-ikd}\right)$, with the effective nearest-neighbor intra- and inter-unit-cell hoppings, respectively, $t_{ab}$ and $t_{ba}$, as shown in Fig.~\ref{fig:model}(b), which are related to the chemical potential, and the GB angle. On the other hand, the intra-sublattice hopping $\tilde{\Delta}$ depends on the pairing amplitude and the geometric details of the superlattice. Finally, $\tilde{V}(k)$, is an effective spin-orbit coupling term induced by the exchange field, as shown explicitly in Supplementary Note 4.     

In the absence of the exchange field, the spectrum of the GB SC reads as 
\begin{align}
E(k) = \pm \sqrt{4\tilde{\Delta}^2\sin^2kd + |W(k)|^2}.
\end{align}
The diagonal part of the Hamiltonian appears due to the hybridization of neighboring edge dislocations on the same sublattice, giving rise to a characteristic sinusoidal dispersion. Furthermore, the hopping term $W(k)$ yields the interlattice hybridization, which for equidistant, thereby mirror-symmetric GB ($l = d/2$ and $t_{ab} = t_{ba}$) reduces to $W(k) = 2t_{ab}\cos (kd/2)$, yielding a node at the $M$ point ($k=\pi/d$) in the superlattice BZ. When the edge dislocations feature a relative slip, as in \cref{fig:bsmu}(b) where $l = 5d/12$, the mirror symmetry is broken, and the SC is gapped. 

With the GB SC being gapped out due to the slip, we now turn on the exchange field, which, as we show in the following,  is crucial for inducing nontrivial topology on the grain boundary. The effective 1D Hamiltonian in Eq.~\eqref{eq:effHam} has particle-hole symmetry, implying that it features two topological classes distinguished by a $\mathbb{Z}_2$ topological invariant, which is determined by the Pfaffian  index~\cite{Kitaev2001}. For the exchange field oriented along the GB line ($\theta=\pi/2$), as we explicitly show in Supplementary Note~5, the GB defect hosts a topologically nontrivial SC, with a pair of localized MZMs,  when 
\begin{equation}
t_{\rm min} < h\cos\alpha < t_{\rm max},
\end{equation}
with $t_{\rm min(max)}={\rm min(max)}\{|t_{ab}-t_{ba}|,|t_{ab}+t_{ba}|\}$, corresponding to  the minimum (maximum) gap along the GB for ${\bm h}=0$. As such, the effective hoppings $t_{ab},t_{ba}$ are highly sensitive to the microscopic details of the system. Most importantly, the two MZMs are protected by an emergent antiunitary symmetry $U=(\sigma_2\otimes\sigma_0)\,{\rm K}$ of the effective 1D GB Hamiltonian in Eq.~\eqref{eq:effHam}, with $U^2=-1$ and ${\rm K}$ as complex conjugation,  implying the orthogonality of the two MZMs. See Supplementary Note~6  for details.

To demonstrate that the effective model \eqref{eq:effHam} can indeed capture topological features of the numerically observed emergent GB SC, we use numerically obtained value of the critical magnetic field for the topological transition $h_{\rm crit}$ from Fig.~\ref{fig:bsh}(a), to infer values of the critical angles $\theta_{c,1}$ and $\theta_{c,2}$ from the effective model. The value of the critical field from Fig.~\ref{fig:bsh}(a), as shown in Supplementary Note~5, yields  $\theta_{c,1}\simeq57^\circ=1.00$ and $\theta_{c,2}\simeq123^\circ=2.14$ radians, very close to the numerically found values, see Fig.~\ref{fig:bsh}(b). Finally, the computed topological invariant of the GB SC for the exchange field along the GB as its magnitude increases, ~\cref{fig:bsh}(c), and as the exchange field rotates, ~\cref{fig:bsh}(d), confirm the topological nature of the GB SC and the concomitant MZMs.

We now further support the robustness of 
the cohabitating Majorana modes at the top of the GB defect, by introducing disorder in the system's bulk,  modeled by a random variation in the chemical potential of up to $10\%$, making sure to explicitly break all crystal symmetries (including  any mirror symmetry). As a result, we find that the two GB and two parent MZMs remain unaffected, i.e. they are still  zero energy states, and localized at the end of the GB and the corners, respectively, thus confirming their robustness against the nonmagnetic disorder. For the details, consult Supplementary Note 7 and Supplementary Figure 4.

{\bf Experimental feasibility.} We now discuss candidate platforms for the experimental realization of our proposal. The key condition is that the parent topological SC should be translationally active, with several possible candidates in this respect. First, the signatures of a  topological SC with Fermi surface away from the $\Gamma-$point have been already observed in Fe-based compounds, for instance, in crystalline domain walls in FeSe$_{0.45}$Te$_{0.55}$ ~\cite{Zhenyu-Wang-2020}. Furthermore, the GB defects can be manipulated in some of the Fe-based superconducting materials~\cite{Katase2011}. Second, quantum wells of doped topological crystalline insulators (TCIs), e.g. SnTe, a paradigmatic TCI, features Fermi pockets close to the $L-$points in the BZ~\cite{Hsieh-NatComm2012}, with recently reported evidence of an unconventional bulk superconductivity when In-doped~\cite{wang-2022}. In addition, the tunability of the GB defects in this material~\cite{Wu-AppliedEnergy2019},  should facilitate not only the realization of the GB SC chain, but can also be used to manipulate the  MZMs for braiding operations.  Finally, designer materials, assembled by scanning tunneling microscopy, offer a realizable platform hosting a 2D translationally-active topological SC ~\cite{Kezilebieke2020}, which through the defect engineering should be beneficial for the realization of the GB Majorana chain.

Irrespective of the concrete platform, as a first step, a pair of dislocation modes should be manipulated to map out the hopping parameters of the effective model in terms of the  magnitude and the angle of the exchange field, for a fixed set of parameters in the parent SC. This mapping then can be used for the manipulation of the GB superlattice in terms of the lattice constant and the slip, so that the Majorana chain is in the topological regime, while the parent SC remains gapped and topological.

 \noindent\\
 {\bf Conclusions}\\
 
We here show that a GB defect in a parent topological SC  can host symmetry and topologically protected MZMs, tunable by the in-plane magnetic (exchange) field, which are localized at the defect's end.
As such, the modes may be detectable by local probes, e.g. scanning tunneling microscopy. Furthermore, we expect that the manipulation of the defect's geometry  provides the control necessary for achieving information storage and  braiding of the MZMs, which may be important for the quantum technology applications. In particular, the  location of the zero modes may be controlled and manipulated by the migration of the  GB defects through the collective glide motion of the elementary dislocations~\cite{Azizi2014-rs}, which is a
 prospect that we will investigate in future. We also point out that the control over the gap of the GB SC  by rotating the exchange  field may be relevant for  applications in spintronics.

Our work opens up a new, to the best of our knowledge, direction in the search for defect-based platforms for
quantum information technology. As such, it will also spur research efforts toward the proposals and possible realization of such platforms  in three spatial dimensions, where a planar GB defect should host propagating edge Majorana modes. Finally, our findings are expected to help further understand the role of lattice defects and   establish new probing setups for topological SCs. 

 \noindent\\
{\bf Methods}\\

We consider the GB defect, with the geometry shown in Fig.~\ref{fig:model} (see also Supplementary Figure 1), in a parent translatioanlly-active topological SC described by the Hamiltonian in Eq.~\eqref{eq:model}. We first numerically implement this Hamiltonian on a real-space square lattice with the GB defect and perform the numerical analysis using the Kwant code, see also Supplementary Note 1 for additional details of the model. After including the dislocation slip and the exchange field, using this procedure,  we numerically demonstrate the occurrence of the topological phase transition for the GB SC and the emergence of the MZMs at the defect's end by following the evolution of the emergent GB band-structure and the LDOS when the magnitude of the magnetic field is tuned  while the field  is oriented along the defect (Fig.~\ref{fig:bsmu}). Additional results  supporting this scenario  are presented in Supplmentary Note~2, Supplementary Figures 2 and 3. To further corroborate the topological nature of this transition and the emergence of the MZMs, we identify the closest-to-zero energy modes at the two side of the transition and show that two additional modes appear in the GB SC (Fig.~\ref{fig:spectrum}).  Finally,  we compute  the critical strength of the exchange field corresponding to the topological phase transition of the GB SC when the field is aligned with the GB defect, see Fig.~\ref{fig:bsh}(a). We also find the critical angle of the topological phase transition for a fixed magnitude of the exchange field, shown in Fig.~\ref{fig:bsh}(b). 

These numerical results are supported by an effective model obtained by projecting the parent Hamiltonian in Eq.~\eqref{eq:model} on the subspace formed by the Majorana
modes at the dislocation defects, representing the Wannier basis for  the GB superlattice, yielding the effective  Hamiltonian given by Eq.~\eqref{eq:effHam}. The Wannier states are  obtained as the zero-modes of the parent Hamiltonian with a single dislocation defect, as shown in Supplementary Note 3, while the derivation of the effective GB SC model is presented in Supplementary Note~4. 

Topological and symmetry analysis of the the  effective model in Eq.~\eqref{eq:effHam} is performed as follows.

\begin{enumerate}[(i)]

\item We compute $\mathbb{Z}_2$ topological invariant of the GB SC for the exchange field along the GB as its magnitude increases, using the standard Pfaffian formulation, as discussed in Supplementary Note~5, with the results shown in ~\cref{fig:bsh}(c). Additionally, we follow the evolution of the topological invariant as the exchange field rotates, ~\cref{fig:bsh}(d), keeping its strength fixed, which further corroborates the topological nature of the GB SC. 

\item To show the additional symmetry protection of the GB SC phase, we rewrite the effective Hamiltonian~\eqref{eq:effHam} in the form that allows  for identification of the protecting antiunitary particle-hole symmetry, and the symmetry operator is given in Supplementary Note 6, Eq.~(S53). The stability of the GB SC  with respect to weak chemical-potential disorder is furthermore numerically demonstrated in Supplementary Note 7, with the results shown in Supplementary Figure 4.  
\end{enumerate}

Additional technical details  are discussed in the Supplementary Information.

\noindent\\
{\bf Data Availability}\\
The data that support the plots within this paper and other findings of this study are
available from the corresponding author upon reasonable request.

\noindent\\
{\bf Code Availability}\\
The software code used in producing the results of the main text and the supplementary materials is available at \url{https://doi.org/10.5281/zenodo.8210977}.

\noindent\\
{\bf Acknowledgments}\\
The authors acknowledge support
of the Swedish Research Council (Grant No. VR 2019-04735 of V.J.). Nordita is partially supported by Nordforsk. 
V. J. is thankful to Bitan Roy for useful discussions. 

\noindent\\
{\bf Author Contributions}\\
M.A. performed all the numerical calculations. M.A  and V. J. performed analytical calculations and wrote the manuscript. V. J. conceived the idea and structured the project.

\noindent\\
{\bf Competing interests}\\
The authors declare no conflicts of interest.

%\printbibliography[heading=myheading]
\bibliographystyle{naturemag}
\bibliography{references}  

\end{document}